\documentclass[12pt,preprint]{aastex}
\shorttitle{A PULSATING ULTRACOOL DWARF}
\shortauthors{Hallinan et al.}

\begin{document}


\title{PERIODIC BURSTS OF COHERENT RADIO EMISSION FROM AN ULTRACOOL DWARF}


\author{G. Hallinan\altaffilmark{1}, S. Bourke\altaffilmark{1}, C. Lane\altaffilmark{1}, A. Antonova\altaffilmark{2}, R. T. Zavala\altaffilmark{3}, W. F. Brisken\altaffilmark{4}, R.P. Boyle\altaffilmark{5}, F. J. Vrba\altaffilmark{3}, J.G. Doyle\altaffilmark{2} and A. Golden\altaffilmark{1}}

\altaffiltext{1}{Computational Astrophysics Laboratory, I.T. Building, National University of Ireland, Galway, Ireland; gregg@it.nuigalway.ie, stephen@it.nuigalway.ie, c.lane2@nuigalway.ie, agolden@it.nuigalway.ie}

\altaffiltext{2}{Armagh Observatory, College Hill, Armagh BT61 9DG, N. Ireland; tan@arm.ac.uk, jgd@arm.ac.uk}

\altaffiltext{3}{United States Naval Observatory, Flagstaff Station, 10391 West Naval Observatory Road, Flagstaff, AZ 86001; bzavala@nofs.navy.mil, fjv@nofs.navy.mil}

\altaffiltext{4}{National Radio Astronomy Observatory, P.O. Box O, Socorro, NM 87801; wbrisken@aoc.nrao.edu}

\altaffiltext{5}{Vatican Observatory Research Group, Steward Observatory, University of Arizona, Tucson, AZ 85721; boyle@ricci.as.arizona.edu}


%\altaffiltext{1}{Visiting Astronomer, Cerro Tololo Inter-American Observatory.
%CTIO is operated by AURA, Inc.\ under contract to the National Science
%\altaffiltext{2}{Society of Fellows, Harvard University.}
%\altaffiltext{3}{present address: Center for Astrophysics,
 %   60 Garden Street, Cambridge, MA 02138}
%altaffiltext{4}{Visiting Programmer, Space Telescope Science Institute}
%\altaffiltext{5}{Patron, Alonso's Bar and Grill}


\begin{abstract}

We report the detection of periodic ($p = 1.96$ hours) bursts of extremely bright, 100\% circularly polarized, coherent radio emission from the M9 dwarf TVLM 513-46546. Simultaneous photometric monitoring observations have established this periodicity to be the rotation period of the dwarf. These bursts, which were not present in previous observations of this target, confirm that ultracool dwarfs can generate persistent levels of broadband, coherent radio emission, associated with the presence of kG magnetic fields in a large-scale, stable configuration. Compact sources located at the magnetic polar regions produce highly beamed emission generated by the electron cyclotron maser instability, the same mechanism known to generate planetary coherent radio emission in our solar system. The narrow beams of radiation pass our line of sight as the dwarf rotates, producing the associated periodic bursts. The resulting radio light curves are analogous to the periodic light curves associated with pulsar radio emission highlighting TVLM 513-46546 as the prototype of a new class of transient radio source. 
  
\end{abstract}





\section{INTRODUCTION}

\par Ultracool dwarfs are generally defined as those dwarfs with spectral type $\geq$ M7 \citep{kirkpatrick97}, encompassing very low mass stars just above the stellar/substellar boundary and all brown dwarfs below it. Surprisingly, despite their low bolometric luminosity, a number of ultracool dwarfs have been detected as relatively bright sources at radio frequencies \citep{berger01,berger02,burgasser05,berger06,phan-bao07}. Initially, the detection of variable, broadband radio emission with low net circular polarization suggested incoherent gyrosynchrotron radiation due to a population of electrons spiraling in the magnetic field of the ultracool dwarf \citep{berger05,osten06}. This is the mechanism by which higher mass dwarf stars are thought to generate persistent levels of radio emission. However, such stars are surrounded by high temperature chromospheres and coronae and the levels of radio emission they produce have been found to be tightly correlated to their coronal X-ray luminosities \citep{gudel93, benz94}. Ultracool dwarfs, on the other hand, are thought to possess cooler, more neutral atmospheres, as evidenced by their low H$\alpha$ and X-ray luminosities \citep{mohanty03,stelzer06}, and yet can produce high levels of radio emission, many orders of magnitude brighter than predicted by this relationship.

\par \citet{berger05} reported a periodicity of $\sim 3$ hours in the radio emission detected from one such ultracool dwarf 2MASS J00361617+1821104 citing orbital motion of a companion, rotation of the dwarf or periodic magnetic reconnection as possible causes of the observed behaviour. \citet{hallinan06} detected a periodicity in the radio emission from a second ultracool dwarf, TVLM 513-46546 (hereafter TVLM 513) consistent with the putative rotation period of the dwarf derived from $v$ sin $i$ data, thereby favouring rotation as the cause of the observed periodicities for both dwarfs. Based on this assumption, the periodic variability can be explained by either (a) an anisotropically beamed radio emission mechanism or (b) the occultation of very high brightness temperature compact regions in the magnetosphere of the dwarf. Both scenarios were deemed inconsistent with isotropic gyrosynchrotron emission from a large extended corona. It was postulated that the radio emission from ultracool dwarfs may instead be due to coherent electron cyclotron maser emission generated in the low density regions above the poles of a large-scale magnetic field \citep{hallinan06}, the same mechanism producing the coherent radiation detected from the magnetized planets in our solar system \citep{zarka98, ergun00}. However, studies of the electron cyclotron maser instability operating in planetary magnetospheres have reported intrinsically narrow banded, 100\% polarized emission. The radio emission from TVLM 513, on the other hand, was found to be broadband with net circular polarization $\lesssim 13\%$ reaching a periodic maximum of $\sim 40\%$, properties more consistent with the incoherent gyrosynchrotron process. We have conducted further deep pointings of TVLM 513 in an effort to unambiguously distinguish between incoherent gyrosynchrotron emission and coherent electron cyclotron maser emission. The periodicity reported by \citet{hallinan06} was once again present, but the dwarf was found to be in a much more active state than in previous epochs, producing periodic bursts of extremely bright 100\% circularly polarized emission. These observations provide direct confirmation that ultracool dwarfs can produce persistent levels of broadband, 100\% polarized, coherent electron cyclotron maser emission.

\section{OBSERVATIONS}

\par We used the Very Large Array (VLA) \footnote[1]{The VLA is operated by the National Radio Astronomy Observatory, a facility of the National Science Foundation operated under cooperative agreement by Associated Universities, Inc.} to observe the M9 dwarf TVLM 513-46546 for $\sim 10$ hours at a frequency of 8.44 GHz on 2006 May 20 and for $\sim 10$ hours at a frequency of 4.88 GHz on 2006 May 21. Data reduction was carried out with the Astronomical Image Processing System (AIPS) software package. The light curve of the 8.44 GHz radio emission is characterized by periodic bursts of both left and right 100\% circularly polarized emission, the brightest periodically reaching flux values of $\sim 5$ mJy, which are stable in phase for the duration of the observation (Fig. 1). Similar, albeit less bright (up to $\sim 1$ mJy), periodic bursts are also present in the 4.88 GHz light curve. The periodicity of 1.96 hours was also retrieved in I band photometric monitoring observations of TVLM 513 using telescopes at the US Naval Observatory (USNO), Flagstaff Station and the Vatican Observatory (VO), Arizona confirming that the periodicity present in the radio emission from TVLM 513 is due to the rotation of the dwarf (Fig. 2). Therefore, although the bright, polarized bursts present in the radio data appear to be transient flares, they are produced by the strong beaming of persistent radio emission together with the rapid rotation of the dwarf. For the limiting case of perfectly beamed emission and assuming a source $< 0.5$R from the surface, where R is the radius of the dwarf, we can use the duration of the bursts to geometrically constrain the size of the emitting source region to be $< 0.22$R. Using this value, we can establish the brightness temperature of the radio emission to be $> 2.4 \times 10^{11}$ K, which, together with the 100\% circular polarization and extremely narrow beaming, provides conclusive confirmation of its coherent nature. 

\par The phase folded light curves of the radio emission detected from TVLM 513 at 8.44 and 4.88 GHz are shown in Fig. 3. The flux detected over the duration of the observation was $0.464 \pm 0.009$ mJy and $0.368 \pm 0.011$ mJy at 8.44 and 4.88 GHz respectively. These values are similar to those reported previously for this dwarf by \citet{hallinan06} of $0.405 \pm 0.018$ mJy and $0.396 \pm 0.016$ mJy at 8.44 and 4.88 GHz respectively. However, the periodic bursts, particularly those detected at 8.44 GHz which are an order of magnitude brighter than the mean flux values, were not detected in the observations of \citet{hallinan06}, indicating that activity levels in the compact regions producing the bursts have increased greatly in the intervening $\sim 16$ months between the two epochs. Although the periodic bursts are 100\% circularly polarized, the interpulse emission at both frequencies is largely unpolarized. This unpolarized component does not display the high degree of periodic variation observed for the 100\% circularly component of the emission indicating much lower intrinsic beaming of the emission. 

\par Although periodic bursts are detected at 4.88 and 8.44 GHz, the phase folded light curves suggest narrowband structure across this frequency range, for example the brightest burst in the 8.44 GHz phase folded light curve does not appear to have a counterpart in the 4.88 GHz light curve. Although there is a burst at a similar phase in the 4.88 GHz light curve, it is left circularly polarized and thus probably originates in a region of opposite magnetic polarity to the right circularly polarized 8.44 GHz burst. It should be noted that direct correlation of the radio emission at both frequencies is subject to the assumption that the periodic bursts were stable in phase and amplitude over the duration of the two night monitoring campaign, as data were not obtained at both frequencies simultaneously. It is notable that the phase folded light curve of the 4.88 GHz emission is characterized by two highly polarized bursts per period of rotation of the dwarf, separated by $\sim 0.5$ phase. This is consistent with emission from the same source region emitted perpendicular to the local magnetic field and hence detected twice per period of rotation. In contrast, the bursts observed at 8.44 GHz are only detected once per rotation and hence may be obscured on one side by optically thick plasma or indeed the stellar surface. 

\section{THE ELECTRON CYCLOTRON MASER}

\par The characteristics of the periodic bursts, which include 100\% circular polarization, high brightness temperature and very narrow beaming, are all consistent with coherent emission generated via the electron cyclotron maser instability \citep{treumann06}. As discussed by \citet{hallinan06}, the unpolarized component of the radio emission detected from TVLM 513 may also be produced by the coherent electron cyclotron maser mechanism, but incoherent gyrosynchrotron or synchrotron emission cannot be ruled out without further multifrequency observations to separately characterize the 100\% circularly polarized bursts and the unpolarized component of the emission. 

\par The electron cyclotron maser instability is the mechanism deemed responsible for the coherent radio emission detected from the magnetized planets in our Solar System \citep{zarka98,ergun00}, and has also been proposed as a source of certain classes of solar and stellar bursts. The bulk of planetary electron cyclotron maser emission is confined to high magnetic latitudes and in-situ measurements of Earth's auroral kilometric radiation (AKR) have provided the deepest insights in to the nature of the radiation mechanism. AKR is associated with the presence of magnetic field-aligned electric fields at high magnetic latitudes that (1) evacuate thermal plasma along open field lines enabling the formation of density cavities where the electron cyclotron frequency can greatly exceed the plasma frequency, $\nu_c \gg \nu_p$, a prerequisite for the generation of electron cyclotron maser emission and (2) accelerate electrons into these density cavities that adiabatically evolve along field lines of increasing strength to form an unstable `horseshoe' or `shell' distribution \citep{pritchett84,pritchett85,winglee86}, providing the free energy to power the maser emission. Emission in the R-X mode perpendicular to the magnetic field in the source region is favoured, resulting in emission of opposite polarity from magnetic polar regions of opposite polarity. A similar model was proposed by \citet{hallinan06} to account for the radio emission detected from ultracool dwarfs, now confirmed to be the case with the detection of periodic 100\% circularly polarized bursts from TVLM 513. 

\par The means by which the stable magnetic field aligned electric fields are generated and maintained in the magnetosphere of TVLM 513 remains unclear. We rule out the interaction of the magnetosphere with a stellar wind from a coronally active star, the means by which a large fraction of planetary maser emission is powered, as high resolution adaptive optics observations of TVLM 513 have found no evidence of a physical companion \citep{close03}.  It is also unlikely that a close-in orbital companion is responsible, as is the case for the component of Jupiter's maser emission associated with the Jupiter-Io electrodynamic circuit. The narrow duty cycle bursts detected from TVLM 513 are tightly locked in phase with rotation of the dwarf and hence are not in corotation with an orbital companion. We can assert therefore that the stable electric fields are somehow generated and sustained within the magnetosphere of the ultracool dwarf. It is worth noting that the rapid rotation of Jupiter is also thought to be an efficient source of electron acceleration for a component of the non-Io related maser emission and a similar scenario may be applicable to TVLM 513.

\section{MAGNETIC FIELD DIAGNOSTICS}

\par Electron cyclotron maser radiation is emitted at the electron cyclotron frequency denoted by $\nu_c \approx 2.8 \times 10^6 B$ Hz, where B is the magnetic field strength in the radio emission source region. It has been proposed that emission at the secondary harmonic may dominate where radiation generated at the fundamental electron cyclotron frequency cannot escape due to gyroharmonic absorption by thermal plasma, for example, in the case of maser emission generated at the base of magnetic flux tubes in the solar corona \citep{melrose82}. However, thermal plasma density is expected to be much lower in the density cavities located at the magnetic polar regions of TVLM 513 than at the base of the solar corona, and therefore optical depth at the secondary harmonic layer should also be much smaller than derived for the solar case. \citet{melrose82} also note that growth at the fundamental is extremely efficient limiting the energy available for growth at higher harmonics. Importantly, measurements by the FAST satellite at Earth and the Ulysses probe at Jupiter have directly confirmed that planetary electron cyclotron maser emission is almost entirely generated at the fundamental frequency. Therefore, emission at the fundamental frequency is strongly favoured for the bursts of electron cyclotron maser emission detected from TVLM 513. We can thus derive the field strength at the source of the 8.44 GHz bursts detected from TVLM 513 to be $\sim3$ kG, confirming the estimate of \citet{hallinan06}. These high magnetic field strengths, which can also be attributed to other ultracool dwarfs previously detected at radio frequencies including a number of bona fide brown dwarfs, are of the same order as those confirmed for the most magnetically active dMe flare stars \citep{johnskrull96}. \citet{reiners07} have recently directly measured the magnetic field strengths of a number of ultracool dwarfs confirming that magnetic flux is indeed maintained at the lower end of the main sequence.

\par The extremely high stability in phase of the narrow duty cycle, 100\% circularly polarized bursts detected from TVLM 513, coupled with the detection of bursts at both 4.88 and 8.44 GHz, confirm that these high strength magnetic fields are in a large-scale, stable configuration. This is further supported by the periodic signal detected in the I band photometric data (Fig. 2), which is stable in phase and amplitude for the duration of the 4 night monitoring campaign, consistent with the presence of high strength, fixed magnetic field regions at the stellar photosphere (Lane et al.\ in preparation). Multiple bursts of both left and right 100\% circularly polarized emission, which originate in regions of opposite magnetic polarity, are confined to the same range of phase of rotation of the dwarf ($\sim0.45 - 0.78$, Fig. 3), indicating the probable presence of a dipolar component to this large-scale magnetic field. Periodic bursts are detected when the axis of this dipolar field lies in the plane of the sky, which occurs twice per period of rotation of the dwarf. This model requires the dipolar field to be tilted relative to the rotation axis of the dwarf. 

\section{DISCUSSION AND CONCLUSIONS}

\par Electron cyclotron maser emission generated at the poles of a large-scale magnetic field may be a ubiquitous source of radio emission from planets, brown dwarfs and very low mass stars. \citet{kellett02} and \citet{bingham01} have also proposed that electron cyclotron maser emission originating in the polar regions of a large-scale magnetic field may be a source of radio emission from coronally active stars, particularly dMe flare stars. The recent magnetic mapping of a rapidly rotating M4 dwarf has revealed the presence of a dipolar component, confirming that such stars can indeed possess large-scale, stable magnetic fields \citep{donati06}. In light of the observed behaviour of TVLM 513 we note that many of the coherent bursts detected from stellar sources, particularly dMe flare stars, might not be transient flare events but rather may be due to the beaming of persistent electron cyclotron maser emission coupled with rotation of the star. Periodicity may have gone undetected for higher mass stars as they have much longer periods of rotation than ultracool dwarfs \citep{zapatero06}, making it difficult to directly associate coherent flares with a particular phase of rotation.

\par In light of the sharp drop observed in the levels of H$\alpha$ and X-ray emission for late M, L and T dwarfs, the electron cyclotron maser may prove a vital diagnostic tool for remote sensing of plasma conditions and magnetic field strengths and topologies in the mass gap between planets and stars. Previous radio surveys have been undertaken of a volume limited sample of ultracool dwarfs in the Solar neighborhood, resulting in an overall detection rate of $\sim 10\%$ \citep{berger06}. However, these surveys were conducted solely at a frequency of 8.44 GHz, important when considering that electron cyclotron maser emission is characterized by an upper cut-off frequency associated with the maximum magnetic field strength in the magnetosphere of the dwarf. A detection at 8.44 GHz requires magnetic field strengths of at least 3 kG. Therefore a much larger fraction of these objects with maximum field strengths below this value, particularly lower mass L and T dwarfs, may be detectable sources at lower frequencies. We highlight the recent observation of periodic bursts of coherent emission from an unknown source towards the Galactic Center, GCRT J1745-3009 \citep{hyman05}, at a frequency of 330 MHz, as the possible detection of an as yet unidentified cool, dim brown dwarf or extrasolar planet in the Solar neighborhood. In particular, the period of 1.28 hours observed for the bursts from GCRT J1745-3009 compares favourably with the period of 0.98 hours observed for the bursts detected at 4.88 GHz from TVLM 513, and the bursts have only been detected in $\sim 7\%$ of the time spent on source thus far \citep{hyman07}, indicating long-term activity variation similar to that observed for TVLM 513.


%% Included in this acknowledgments section are examples of the
%% AASTeX hypertext markup commands. Use \url without the optional [HREF]
%% argument when you want to print the url directly in the text. Otherwise,
%% use either \url or \anchor, with the HREF as the first argument and the
%% text to be printed in the second.

\acknowledgments

We gratefully acknowledge the support of the HEA funded Cosmogrid project, the Irish Research Council for Science, Engineering and Technology (IRCSET) and Enterprise Ireland. AG is supported by Science Foundation Ireland (Grant Number 07/RFP/PHYF553). Armagh Observatory is grant-aided by the N. Ireland Dept. of Culture, Arts \& Leisure. We would like to thank Ray Butler, James Mc Donald and Evan Keane for assistance with certain aspects of this manuscript and the referee Manuel G\"udel whose extensive comments significantly improved this manuscript.


%% in parentheses, after the keyword, but they will not be verified.



%\bibliography{ghbib}

\clearpage

\begin{figure}
\epsscale{0.6}
\plotone{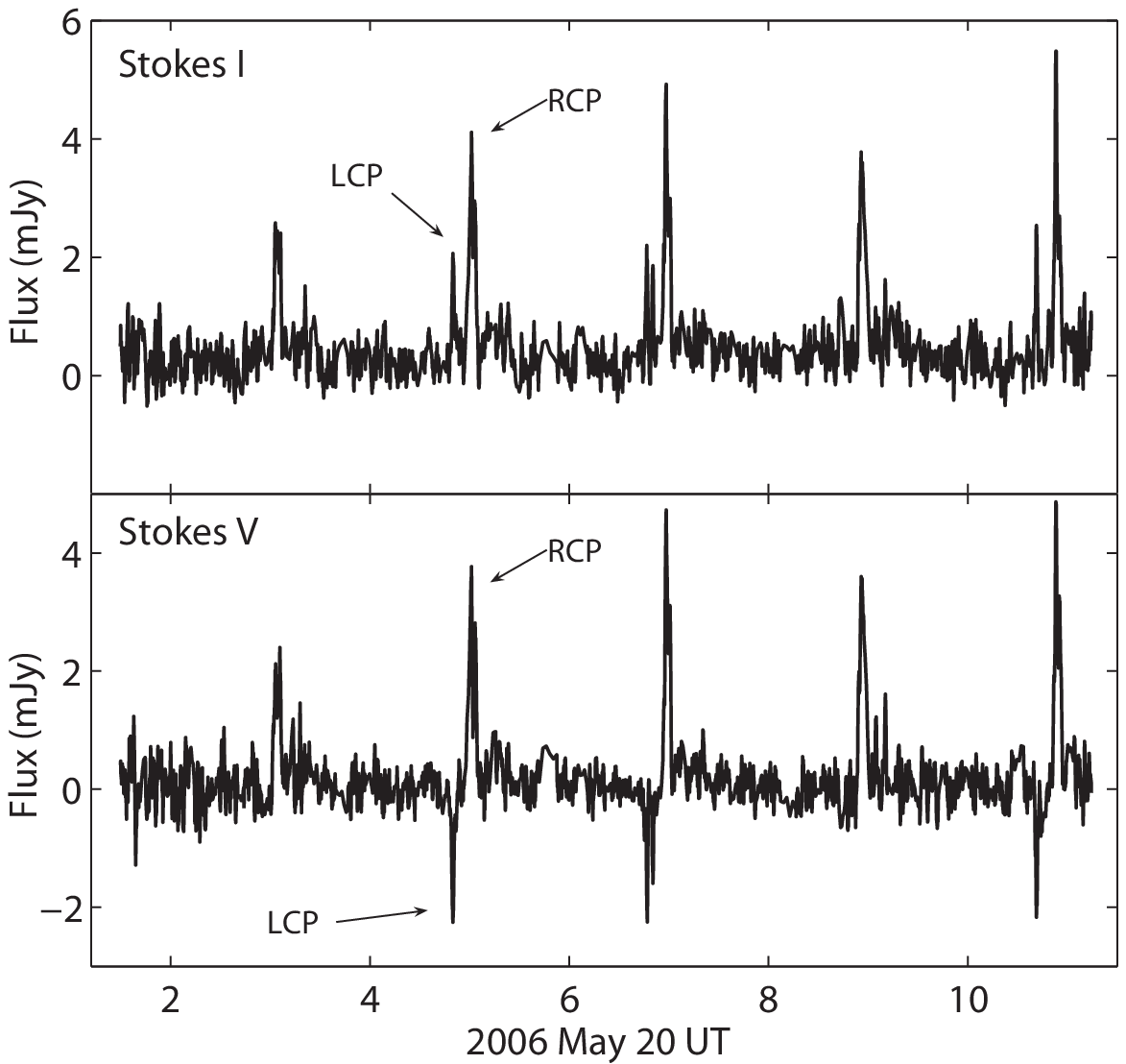}
\caption{The light curves of the total intensity (Stokes I) and the circularly polarized (Stokes V) radio emission detected at 8.44 GHz from TVLM 513. Right circular polarization is represented by positive values and left circular polarization is represented by negative values in the Stokes V light curve. Bursts of both 100\% right circularly polarized emission (an example is highlighted as RCP) and 100\% left circularly polarized emission (an example is highlighted as LCP) are detected with a periodicity of 1.96 hours. The absence of a left circularly polarized burst at approximately 8.7 UT is due to the VLA being pointed at a phase calibration source.}
\end{figure}

\clearpage

\begin{figure}
\epsscale{0.6}
\plotone{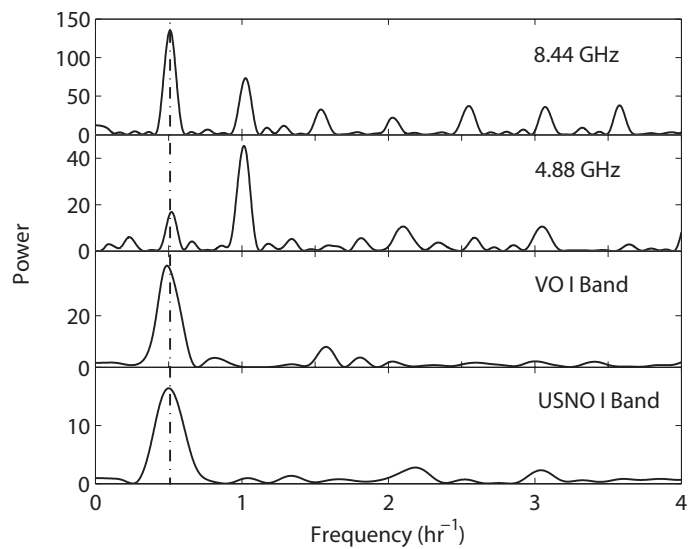}
\caption{Lomb-Scargle periodograms of (a) the 8.44 GHz radio data obtained on May 20 2006 (b) the 4.88 GHz radio data obtained on May 21 2006 (c) the I band photometric monitoring data obtained on May 21 2006 with the Vatican Observatory (VO) 1.8m telescope and (d) the I band photometric monitoring data obtained on May 18 2006 with the US Naval Observatory (USNO) Flagstaff Station 1m telescope. The dashed line corresponds to the putative rotational period of 1.96 hours for TVLM 513. The periodicity present in the radio data is also present in the optical data confirming rotation of the dwarf as the source of this periodicity.}     
\end{figure}

\clearpage

\begin{figure}
\plotone{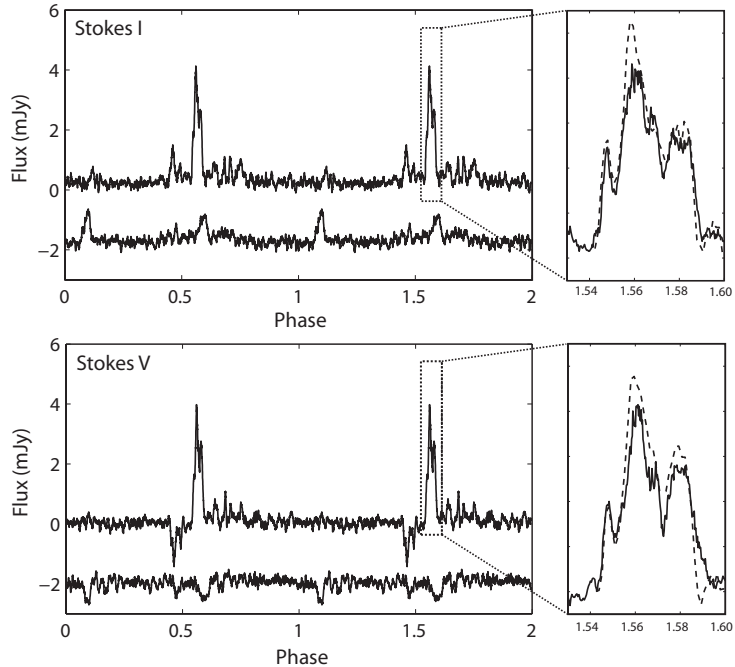}
\caption{Light curves of the total intensity (Stokes I) and circularly polarized (Stokes V) radio emission detected from TVLM 513 at 8.44 and 4.88 GHz (4.88 GHz shifted by -2 mJy) phase folded to a period of 1.958 hours. For the circularly polarized (Stokes V) light curves, right circular polarization is represented by positive values and left circular polarization is represented by negative values. Two periods of each phase folded light curve are shown for clarity. \textit{Inset}: Closer examination of the brightest burst detected at 8.44 GHz reveals further structure, possibly due to the anisotropic morphology of the source region of this burst. The dashed line corresponds to an unfolded light curve of one of the bursts correlated with the phase folded light curve to confirm that this further structure is not a product of the phase folding of the data.}     
\end{figure}

\end{document}